\documentclass[aps,prb,twocolumn,groupedaddress,showpacs]{revtex4-1}

\usepackage{graphicx,color,url,hyperref}

\begin{document}

\title{
  Magnetism trends in
  doped Ce-Cu intermetallics in the vicinity of quantum criticality:\\
  realistic Kondo lattice models based on dynamical mean-field theory
}

\author{
Munehisa~Matsumoto$^{1,2}$
}
\affiliation{
  $^1$Institute for Solid State Physics, University of Tokyo, Kashiwa 277-8581, JAPAN\\
  $^2$Institute of Materials Structure Science,
High Energy Accelerator Research Organization, 1-1 Oho, Tsukuba, Ibaraki 305-0801, Japan
}

\date{\today}

\begin{abstract}
  The quantum critical point (QCP) in the archetypical heavy-fermion compound CeCu$_6$
  doped by Au
  is described, accounting for the localized $4f$-electron of Ce,
  using realistic electronic structure calculations combined with dynamical mean-field theory (DMFT).
  Magnetism trends in Ce(Cu$_{1-\epsilon}$Au$_\epsilon$)$_6$
  ($0<\epsilon\ll 1$)
  are compared with those in
  Co-doped CeCu$_{5}$, which resides on the non-ferromagnetic side of the composition space
  of one of the earliest
  rare-earth permanent magnet compounds, Ce(Co,Cu)$_5$.
  The construction of a realistic Doniach phase diagram shows that
  the system crosses over a magnetic quantum critical point in the Kondo lattice in $0.2<x<0.4$
  of Ce(Cu$_{1-x}$Co$_x$)$_5$.
  Comparison between Au-doped CeCu$_6$ and Co-doped CeCu$_5$ reveals
  that the swept region in the vicinity of QCP for the latter thoroughly covers that of the former.
  The implications of these trends on the coercivity of the bulk rare-earth permanent magnets are discussed.
\end{abstract}


%
%
%

\maketitle

\section{Motivation}
While heavy-fermion (HF) materials and rare earth permanent magnets (REPM's)
have gone through contemporary developments
since the 1960s~\cite{kondo,strnat,nesbitt,tawara,ott,doniach,steglich,sagawa,croat},
apparently little overlap has been identified between the two classes of materials.
One of the obvious reasons
for the absence of mutual interest lies in the difference in the scope
of the working temperatures: HF materials typically
concern low-temperature physics of the order of 10K or even lower
while REPM concerns room temperature at 300K or higher. The other reason
is that the interesting regions in the magnetic phase diagram sit on the opposite sides,
where HF behavior appears around a region where magnetism disappears~\cite{doniach}
while with REPM the obvious interest lies in the middle of a ferromagnetic phase.
In retrospect, several common threads in the developments for HF compounds and REPM's
can be seen: one of the earliest REPM's was Ce(Co,Cu)$_5$~\cite{tawara}
where Cu was added to CeCo$_5$ to implement coercivity, and CeCu$_5$ was
eventually to be identified as an antiferromagnetic Kondo lattice~\cite{bauer_1987,bauer_1991}.

One of the representative HF compounds is CeCu$_6$~\cite{onuki, stewart}
that was discovered almost at the same time as the champion magnet compound Nd$_2$Fe$_{14}$B~\cite{sagawa,croat,rmp_1991}.
While REPM's make a significant part in the most important materials in the upcoming decades
for a sustainable solution of the energy problem
with their utility in traction motors of (hybrid) electric vehicles and power generators,
HF materials might remain to be mostly of academic interests. But we note
that a good permanent magnet is made of a ferromagnetic main-phase
and less ferromagnetic sub-phases.
For the latter compounds in REPM's,
we discuss possible
common physics with HF materials,
namely, magnetic quantum criticality where
magnetism disappears and associated scales
in space-time fluctuations diverge,
and propose
one of the possible
solutions for a practical problem on
how to implement coercivity, which measures
robustness of
the metastable state with magnetization
against externally applied magnetic fields.

Even though the mechanism of bulk
coercivity on the macroscopic scale
in REPM's is not entirely understood,
the overall multiple-scale structure has been clear in that
the intrinsic properties of materials on the microscopic scale of the scale of $O(1)$~nm
is carried over to the macroscopic scale via mesoscopic scale.
Namely, possible scenarios in coercivity
of Nd-Fe-B magnets~\cite{hono_2012,hono_2018} and Sm-Co magnets~\cite{ohashi_2012}
have been so far discussed
as follows.
\paragraph{Nd-Fe-B magnets} Propagating domain walls
around a nucleation center of reversed magnetization are blocked before going too far.
Infiltrated
elemental Nd in the grain-boundary region
that is paramagnetic in the typical operation temperature range of $O(100)$~K
neutralizes inter-granular magnetic couplings among Nd$_2$Fe$_{14}$B grains~\cite{murakami_2014}.
Single-phase Nd$_2$Fe$_{14}$B does not show coercivity at room temperature
and
fabrication of an optimal microstructure on the mesoscopic scale,
with the infiltrated Nd metals
between Nd$_2$Fe$_{14}$B grains, seems to be crucial to observe bulk coercivity.
\paragraph{Sm-Co magnets and Ce analogues} Pinning centers of domain walls
are
distributed over cell-boundary phases made of Sm(Co,Cu)$_5$
which separate hexagonally-shaped cells of Sm$_{2}$(Co,Fe)$_{17}$.
The uniformity of the cell boundary phase~\cite{ohashi_2012,navid_2017} suggests that the pinning intrinsically
happens on the microscopic scale in Sm(Co,Cu)$_5$
which freezes out the magnetization reversal dynamics.
Also for CeCo$_5$, addition of Cu
has been found to help the development of bulk coercivity~\cite{tawara}
without much particular feature in the microstructure, suggesting here again
an intrinsic origin contributing to the bulk coercivity.

Solution of
the overall coercivity problem
takes out-of-equilibrium statistical physics, multi-scale simulations
involving the morphology of the microstructure in the intermetallic materials,
electronic correlation in $4f$-electrons, finite-temperature magnetism
of Fe-based ferromagnets, and magnetic anisotropy,
each of which
by itself
makes a subfield for intensive studies.
Faced with such a seemingly intractable problem, it is important to build up
fundamental understanding step by step.
Therefore, we clarify the magnetism trends around quantum criticality in Ce-Cu intermetallics,
as a part of
$4f$-$3d$ intermetallics that belong to a common thread between HF materials
and REPM, in order to pin-point a possible intrinsic contribution to the coercivity,
specifically via exponentially growing length scales in spatial correlation and characteristic time in the dynamics.

The magnetization in REPM's derives from $3d$-electron ferromagnetism
coming from Fe-group elements and $4f$-electrons in rare-earth elements
provide the uni-axial magnetic anisotropy for the intrinsic origin
of coercivity.
Sub-phases are preferably free from ferromagnetism
to help coercivity e.g. by stopping the propagation of domain walls.
In the practical fabrication of REPM, both of the main-phase compound and other compounds for sub-phases
should come out of a pool of the given set of ingredient elements.
Investigations on non-ferromagnetic materials
that appear in the same composition space as the ferromagnetic material
are of crucial importance for contributing the intrinsic information into
the solution of the coercivity problem.

\begin{figure}
  \begin{center}
\scalebox{0.23}{
\includegraphics{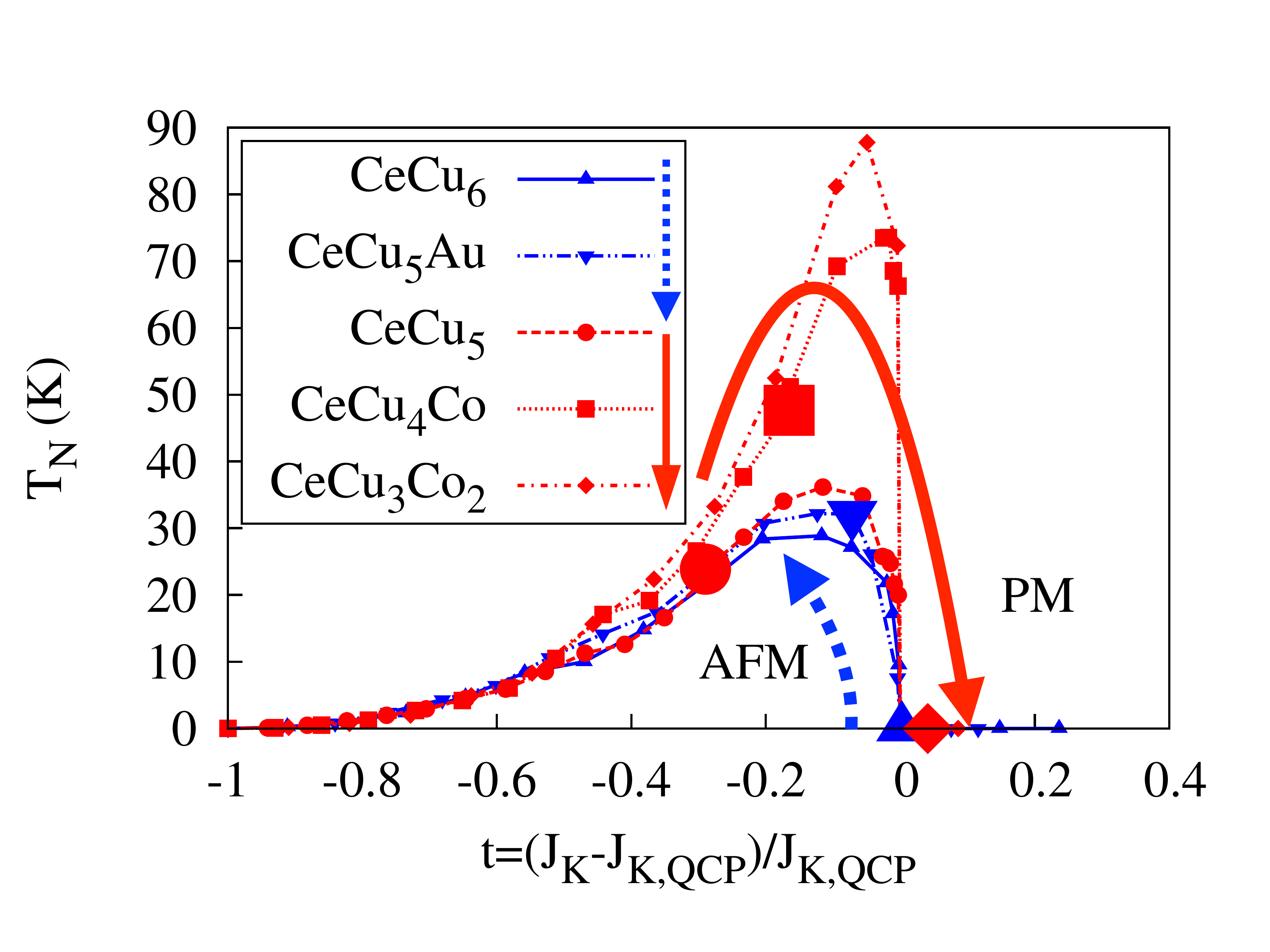}}
\end{center}
\caption{\label{fig::rescaled_Doniach} (Color online) Realistic Doniach phase diagram for the target compounds
  with a rescaled horizontal axis
  to measure an effective distance to the magnetic quantum critical point for each target compound.
  There are the antiferromagnetic (AFM) phase and the Kondo-screened paramagnetic (PM) phase.
  It is seen
  that Co-doped CeCu$_5$ moves from the magnetic side towards the Kondo-screened phase crossing QCP,
  while Au-doped CeCu$_6$ moves to the opposite direction. Arrows are guide for the eye.}
\end{figure}
Thus we investigate the Cu-rich side of the composition space in Ce(Co,Cu)$_5$
and inspect the magnetism trends around the HF compound, CeCu$_5$. It is found
that Co doping into CeCu$_5$ drives the material toward a magnetic quantum critical point (QCP),
to the extent that $3d$-electron ferromagnetism coming from Co does not dominate,
which seems to be the case experimentally~\cite{girodin_1985} when the concentration of Co is below 40\%.
It has also been known that Au-doped CeCu$_6$ goes into quantum criticality~\cite{1994_cecu5.9au0.1,2000_cecu5au},
a trend which is reproduced in the same simulation framework.
With CeCu$_6$ as
one of the most representative HF materials, experimental measurements and
theoretical developments~\cite{si_2001,coleman_2001,rmp_2020}
have been extensively done. Our finding basically reproduces what has already been agreed on the location of
magnetic QCP, but the spirit of our
microscopic description may not entirely be the same as some of the past theoretical works~\cite{si_2001,coleman_2001,rmp_2020}.
Our description should be more consistent with even older works~\cite{anderson_1961}
in the fundamental spirit with the proper incorporation
of realistic energy scales based on electronic structure calculations. We may fail in catching some subtlety
specific to Au-doped CeCu$_6$, but our approach should be suited rather for general purposes in providing an overview over
intrinsic magnetism of $f$-$d$ intermetallics to extract the common physics therein.

We set up a realistic Kondo lattice model~\cite{mm_2009, mm_2010}
for these cases and see the followings: 1) CeCu$_6$ sits very close to the QCP,
2) Au-induced QCP
can also be described on the basis of
a conventional Kondo lattice model
as downfolded from realistic electronic structure data
featuring
localized $4f$-electrons,
at least concerning the relative location of QCP,
without invoking valence fluctuations~\cite{2018}
or the specialized Kondo-Heisenberg model
to describe local quantum criticality~\cite{si_2001,coleman_2001,rmp_2020},
in contrast to some of those previous developments~\cite{si_2001,coleman_2001,2018} for Au-doped CeCu$_6$,
and 3) Co-doping in CeCu$_5$ drives
the material toward the QCP in the opposite direction as Au-doping does in CeCu$_6$.
The main results are summarized in Fig.~\ref{fig::rescaled_Doniach} where the Au-doped CeCu$_6$
and Co-doped CeCu$_5$ are located around a magnetic QCP following a rescaled
realistic Doniach phase diagram~\cite{doniach,mm_2009,mm_2010}.

The rest of the paper is organized as follows. In the next section
we describe our methods~\cite{mm_2009,mm_2010} as specifically applied
to the target materials: pristine CeCu$_6$, CeCu$_5$, and doped cases.
In Sec.~\ref{sec::results} magnetism trends in the target materials
are clarified. In Sec.~\ref{sec::discussions} several issues remaining
in the present descriptions and possible implications from HF physics
on the intrinsic part
of the solution of the coercivity problem of REPM are discussed.
The final section is devoted to the conclusions and outlook.

\section{Methods and target materials}
\label{sec::methods}

We combine {\it ab initio}
electronic structure calculations
on the basis of the full-potential linear muffin-tin orbital method~\cite{andersen_1975,sergey_1996}
and dynamical mean-field theory (DMFT)
for a Kondo lattice model with
well-localized $4f$-electrons~\cite{otsuki_2007,otsuki_2009,otsuki_2009_II},
to construct a Doniach phase diagram~\cite{doniach} adapted
for a given target material to identify an effective distance of the material to a magnetic quantum critical point.
Electronic structure calculations follow
density functional theory (DFT)~\cite{hohenberg_1964,kohn_1965}
within the local density approximation (LDA)~\cite{kohn_1965,vosko_1980}.
Our realistic simulation framework can be regarded as a simplified approach
inspired by
LDA+DMFT~\cite{anisimov_1997,gabi_2006}, where
electronic structure calculations describing the relatively high-energy scales and
a solution of the embedded
impurity problem in the lowermost energy scales are bridged:
here a realistic Kondo lattice model is downfolded~\cite{imada_2010}
from the electronic structure calculations for
Ce-based compounds with well localized $4f$-electrons~\cite{mm_2009,mm_2010}.

More specifically, our computational framework is made of the following two steps:
\begin{enumerate}
\item For a given target material, 
LDA+Hubbard-I~\cite{hubbard_1963-1964,gabi_2006} is done
to extract hybridization between localized $4f$-electrons and conduction electrons,
$-\Im\Delta(\omega)/\pi$ as a function of energy $\hbar\omega$ around the Fermi level.
Position of the local $4f$-electron level below the Fermi level is determined as well.
\item A realistic Kondo lattice model (KLM) with the Kondo coupling $J_{\rm K}$
  is defined following the relations~\cite{mm_2009}:
\begin{eqnarray}
J_{\rm K} & = & \left|V\right|^2 
\left
[\frac{1}{|\epsilon_{f}|}+\frac{1}{(\epsilon_{f}+U_{ff}-J_{\rm Hund})}
\right],
\label{eq::sw} \\
\left|V\right|^2 & \equiv & - \frac{1}{\pi}
\int_{-\infty}^{D}d\,\omega \frac{{\rm Tr}\Im\Delta(\omega)}{N_{\rm F}},
\label{eq::hyb}
\end{eqnarray}
which is a realistic adaptation of
Schrieffer-Wolff transformation~\cite{schrieffer_1966}
to map the Anderson model~\cite{anderson_1961} to Kondo model.
Here $U_{ff}$ and $J_{\rm Hund}$ are the Coulomb repulsion energy and an effective
Hund coupling between $4f$ electrons, respectively,
in $(4f)^2$ configuration and $D$ is an
energy cutoff~\cite{mm_2009,imada_2010}
that defines the working energy window for the realistic
Schrieffer-Wolff transformation. The trace in Eq.~(\ref{eq::hyb})
is taken over all $4f$-orbitals and dividing the traced hybridization by
$N_{\rm F}\equiv 14$ gives the strength of hybridization per each orbital.
Experimental information on the local level splittings is incorporated
for the $4f$-electron part. The thus defined KLM
is solved within DMFT using the continuous-time quantum Monte Carlo impurity solver~\cite{otsuki_2007}.
A Doniach phase diagram~\cite{doniach} separating the magnetic phase and paramagnetic phase
is constructed for each of the target materials and the magnetic QCP is located.
\end{enumerate}
The realistic model parameters that appear in Eqs.~(\ref{eq::sw})~and~(\ref{eq::hyb})
are taken on an empirical basis referring to past works~\cite{gabi_2006,cowan},
among which the origin of the on-site Coulomb repulsion energy $U_{ff}=5$~eV
between $4f$-electrons
can be traced partly
back to past electronic structure
calculations~\cite{herbst_1978} and analyses
of photoemission spectroscopy data~\cite{baer_1979}.
Even though one can argue for material-specific data of $U_{ff}$,
here we are more concerned with relative trends among the target materials
within a realistic model with fixed parameters to get an overview over a group
of Ce-based compounds with well localized $4f$-electrons,
rather than pursuing preciseness of each material-specific data point.

Below we describe the details of the overall procedure
one by one,
taking CeCu$_6$ as a representative case,
partly introducing the results.

\subsection{LDA+Hubbard-I}

\begin{table}
  \begin{tabular}{ccc} \hline
    compound & $a~\mbox{(a.u.)}$,$b/a$,$c/a$ &  \\ \hline\hline
    CeCu$_6$ & $a=15.3295$, $b/a=0.62894$, $c/a=1.25271$ & Ref.~\onlinecite{act_cryst_1960}\\
    CeCu$_5$Au & $a=15.5902$, $b/a=0.61624$, $c/a=1.25576$ & Ref.~\onlinecite{act_cryst_b_1993} \\ \hline\hline
    CeCu$_5$ & $a=9.702$, $b/a=1$, $c/a=0.79957$ & Ref.~\onlinecite{girodin_1985} \\ \hline
    CeCu$_{4}$Co & (fixed to be the same as CeCu$_5$) & \\ \hline
    CeCu$_{3}$Co$_{2}$ & (fixed to be the same as CeCu$_5$) & \\ \hline\hline
  \end{tabular}
\caption{\label{table::inputs} Inputs to LDA+Hubbard-I: the lattice constants of each target compound.}
\end{table}
The overall initial input here is the experimental lattice structure.
This is taken from the past experimental literature for pristine CeCu$_6$
in Ref.~\onlinecite{act_cryst_1960} and CeCu$_5$ in Ref.~\onlinecite{girodin_1985},
and also for CeCu$_5$Au in Ref.~\onlinecite{act_cryst_b_1993}
together with the particular
site preference of the dopant atom, Au. Our input lattice constants are summarized in Table~\ref{table::inputs}.
We note that CeCu$_6$
undergoes a
structural phase transition between a high-temperature orthorhombic phase~\cite{act_cryst_1960}
and a low-temperature monoclinic phase~\cite{JPSJ_1986},
while CeCu$_5$Au does not~\cite{PRB_1999}. In order to compare CeCu$_6$ and CeCu$_5$Au on an equal footing
and inspect the relative trends between them
and observing that the lattice distortion introduced by the structure transition seems to be minor~\cite{PRB_1999},
we fix the working lattice structure
of CeCu$_6$ to be the orthorhombic phase and proceed to the downfolding to the realistic Kondo-lattice model.
The internal coordinates of atoms in CeCu$_6$ and CeCu$_5$Au are shown in Table~\ref{table::internal_coordinates_CeCu6}.
\begin{table}
  \begin{tabular}{l}
    (a) \\
    \begin{tabular}{ccc} \hline
      atom & Wyckoff & internal coordinate \\ \hline
      Ce    & $4c$ & $(0.2602, 0.2500, 0.4354)$ \\ \hline
      Cu(1) & $8d$ & $(0.4354, 0.0041, 0.1908)$ \\ \hline
      Cu(2) & $4c$ & $(0.1467, 0.2500, 0.1418)$ \\ \hline
      Cu(3) & $4c$ & $(0.1821, 0.7500, 0.2451)$ \\ \hline
      Cu(4) & $4c$ & $(0.4380, 0.7500, 0.4023)$ \\ \hline
      Cu(5) & $4c$ & $(0.0987, 0.7500, 0.4846)$ \\ \hline
    \end{tabular} \\
    \\
    (b) \\
    \begin{tabular}{ccc} \hline
      atom & Wyckoff & internal coordinate \\ \hline
      Ce    & $4c$ & $(0.26078, 0.2500, 0.43593)$ \\ \hline
      Cu(1) & $8d$ & $(0.43493, 0.0013, 0.18791)$ \\ \hline
      Au    & $4c$ & $(0.14216, 0.2500, 0.13897)$ \\ \hline
      Cu(3) & $4c$ & $(0.18475, 0.7500, 0.25071)$ \\ \hline
      Cu(4) & $4c$ & $(0.44400, 0.7500, 0.39553)$ \\ \hline
      Cu(5) & $4c$ & $(0.09178, 0.7500, 0.48395)$ \\ \hline
    \end{tabular}
  \end{tabular}
  \caption{\label{table::internal_coordinates_CeCu6} Inputs to LDA+Hubbard-I:
    internal coordinates of atoms in
    the orthorhombic (Space Group No.62) unit cell of (a) CeCu$_6$ and (b) CeCu$_5$Au.
    The spatial translation vectors are plainly $(1,0,0)$, $(0,1,0)$, and $(0,0,1)$,
    measured with respect to the lattice constants as the unit length.
    Each atom in the unit cell has been specified with the Wyckoff position
    and the internal coordinate. The input data are
    taken from (a) Ref.~\onlinecite{act_cryst_1960} and (b) Ref.~\onlinecite{act_cryst_b_1993}.
    It is to be noted that all of the atoms
    in each designated Wyckoff position
    contribute four atoms
    to the unit cell made of four formula units,
    except for Cu(1), which contributes eight atoms to the unit cell.
    Au has been selectively put into the Cu(2) site~\cite{act_cryst_b_1993}.}
\end{table}
\begin{table}
    \begin{tabular}{l}
      \begin{tabular}{ccc} \hline
        atom & Wyckoff & internal coordinate \\ \hline
        Ce & $1a$ & $(0, 0, 0)$ \\ \hline
        Cu & $2c$ & $(1/2, 1/(2\sqrt{3}), 0)$ \\
        Cu & $2c$ & $(1/2, -1/(2\sqrt{3}),0)$ \\ \hline
        Cu & $3g$ & $(1/4,-\sqrt{3}/4,1/2)$ \\ 
        (Cu/Co)$^a$ & $3g$ & $(1/4, \sqrt{3}/4, 1/2)$ \\
        (Cu/Co)$^b$ & $3g$ & $(1/2,0,1/2)$ \\ \hline
      \end{tabular}\\
      $^a$Cu for CeCu$_5$ and CeCu$_4$Co / Co for CeCu$_3$Co$_2$\\
      $^b$Cu for CeCu$_5$ / Co for CeCu$_4$Co and CeCu$_3$Co$_2$
    \end{tabular}
    \caption{\label{table::internal_coordinates_CeCu5} Inputs to LDA+Hubbard-I:
      internal coordinates of the atoms
      in the hexagonal (Space Group No. 191) unit cell
      of CeCu$_5$, CeCu$_4$Co, and CeCu$_3$Co$_2$. Here the spatial translation vectors
      are taken as $(-1/2,\sqrt{3}/2,0)$, $(1/2,\sqrt{3}/2,0)$, and $(0,0,1)$.
      In contrast to Table~\ref{table::internal_coordinates_CeCu6},
      all of the constituent atoms in the same sublattice specified with the Wyckoff position
      are explicitly shown since the replacement of atoms
      happens for a selection of the atoms in the Cu($3g$) sublattice.}
\end{table}
For Co-doped CeCu$_5$, various things happen in real experiments starting with the introduction of a ferromagnetic
conduction band coming from Co and lattice shrinkage even before reaching the valence transition on the Co-rich side.
Here in order to simplify the problem and to focus on the magnetism trends concerning the $4f$-electron QCP,
we fix the working lattice to be that of pristine CeCu$_5$ and inspect the effects of replacements of Cu by Co.
Following the site preference of Co for Cu($3g$) site as suggested in Ref.~\onlinecite{uebayashi_2002}
for Cu-substituted YCo$_5$, which we also confirm in separate calculations~\cite{mm_2018},
we replace Cu by Co in the $3g$ sublattice one by one
as shown in Table.~\ref{table::internal_coordinates_CeCu5}
for CeCu$_4$Co and CeCu$_3$Co$_2$.
With this particular set-up, the effects of Co-doping on CeCu$_5$ has been effectively softened in our calculations.
However
we will see that Co-doping on CeCu$_5$
drives the material across the QCP
more effectively than Au-doping does for CeCu$_6$.

\begin{figure}
\scalebox{0.7}{\includegraphics{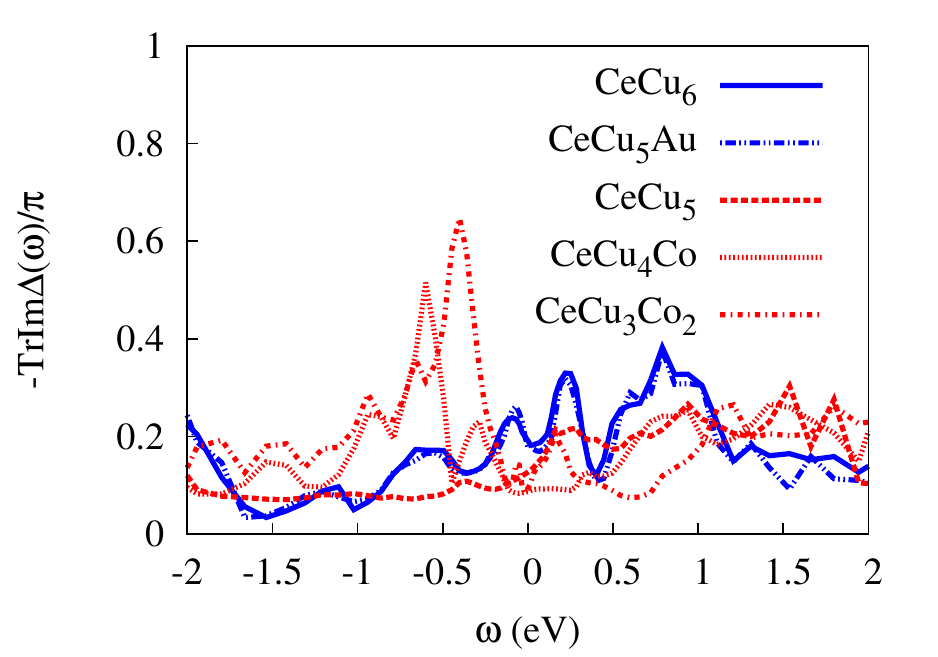}}
\caption{\label{fig::hyb} (Color online) Calculated hybridization function for the target compounds within LDA+Hubbard-I.}
\end{figure}

LDA+Hubbard-I calculations give the hybridization $-\Im\Delta(\omega)/\pi$
and position of the local $4f$-level, $\epsilon_{f}$.
The results for $\epsilon_f$ and $|V|^2$ as defined in Eq.~(\ref{eq::hyb})
are summarized in Table~\ref{table::e_f}. Raw data for $-\Im\Delta(\omega)/\pi$ as traced over
all of the $4f$-orbitals is shown in Fig.~\ref{fig::hyb}.
\begin{table}
\begin{tabular}{ccc} \hline
compound & $\epsilon_f~\mbox{(eV)}$ & $|V|^2$\\ \hline\hline
CeCu$_6$ &  $-1.61$ & $0.172967$\\ \hline
CeCu$_5$Au &  $-1.81$ & $0.159501$\\ \hline\hline
CeCu$_5$ &  $-2.02$ & $0.157148$\\ \hline
CeCu$_{4}$Co &  $-1.99$ & $0.156348$ \\ \hline
CeCu$_{3}$Co$_{2}$ &  $-1.72$ & $0.157907$ \\ \hline\hline
\end{tabular}
\caption{\label{table::e_f}
Outputs of LDA+Hubbard-I: calculated position of localized $4f$-electron level,
$\epsilon_f$, where the offset is taken at the Fermi level, is shown in the second column
for each target compound.
In the third column, the integrated
hybridization as defined in Eq.~(\ref{eq::hyb}) is shown.}
\end{table}

\subsection{DMFT for the realistic Kondo lattice model} 

Following Ref.~\onlinecite{mm_2009}, the hybridization function between the localized
$4f$-orbital in Ce and conduction electron band
defines the material-specific KLM.
Here we describe the details of the Kondo impurity problem
embedded in the KLM within DMFT~\cite{georges_1996} where
we use the continuous-time quantum Monte Carlo solver~\cite{werner_2006}
for the Kondo impurity problem~\cite{otsuki_2007}.

\begin{figure}
\scalebox{0.7}{\includegraphics{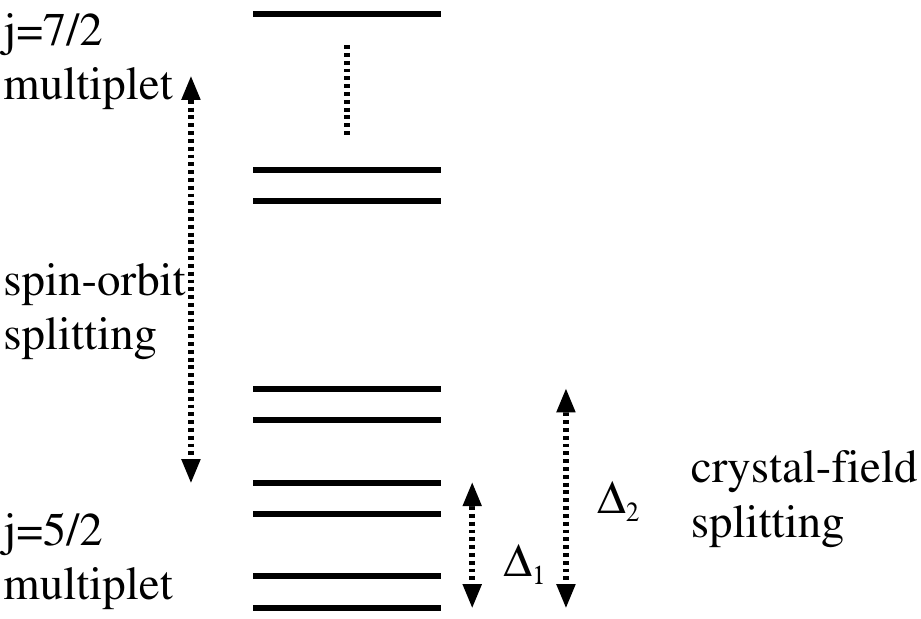}}
\caption{\label{fig::level_splittings} Schematic picture
for local-level splitting caused by spin-orbit interaction
and crystal fields.}
\end{figure}
\begin{table}
\begin{tabular}{ccc} \hline
compound & crystal-field splittings & \\ \hline
CeCu$_6$ & $\Delta_{1}=7~\mbox{meV}$, $\Delta_{2}=13~\mbox{meV}$ & Ref.~\onlinecite{cecu6_2007} \\ \hline
CeCu$_5$ & $\Delta_{1}\simeq\Delta_{2}=17~\mbox{meV}$ & Ref.~\onlinecite{jmmm_1990} \\ \hline
\end{tabular}
\caption{\label{table::KLM_the_inputs} Input crystal-field splittings
following past neutron scattering experiments.}
\end{table}
In the impurity problem embedded in DMFT we incorporate the realistic crystal-field and spin-orbit level splittings in the
local $4f$-orbital of Ce.
The local $4f$-electron level scheme is shown in Fig.~\ref{fig::level_splittings}.
For CeCu$_6$ and hexagonal CeCu$_5$, it is known that the crystal structure splits
the $j=5/2$ multiplets into three doublets, separated by $\Delta_{1}$~(meV) between the lowest
doublet and the second-lowest doublet, and $\Delta_{2}$~(meV)
between the lowest doublet and the third-lowest doublet. Crystal-field splittings have been
taken from the past neutron scattering experiments as summarized in Table~\ref{table::KLM_the_inputs}.
We set the level splitting
between $j=5/2$ and $j=7/2$ multiplets due to spin-orbit interaction
to be $\Delta_{\rm spin-orbit}=0.3$~(eV) referring to the standard situation
in Ce-based HF compounds~\cite{settai_2007}.

The input obtained with LDA+Hubbard-I
to our Kondo problem is shown in Fig.~\ref{fig::hyb}.
The Kondo coupling $J_{\rm K}$
via a realistic variant~\cite{mm_2009} of the Schrieffer-Wolff transformation~\cite{schrieffer_1966}
is defined as in Eqs.~(\ref{eq::sw})~and~(\ref{eq::hyb}).
There $D$ was the band cutoff that is
set to be equal to the Coulomb repulsion $U_{ff}=5$~(eV),
and $J_{\rm Hund}$ is the effective Hund
coupling in the $f^2$ multiplet to which the second term of Eq.~(\ref{eq::sw})
describes the virtual
excitation from the $(4f)^1$ ground state.

We sweep $J_{\rm Hund}$ to locate the QCP on a Doniach phase diagram and also to pick
up the realistic data point at $J_{\rm Hund}=1$~(eV).
This particular choice of the Hund coupling in the virtually excited state $(4f)^2$
has been motivated~\cite{muehlschlegel_1968,cowan} by the typical intra-shell
direct exchange coupling of $O(1)$~eV
and an overall magnetism trend
in CeM$_2$Si$_2$ (M=Au, Ag, Pd, Rh, Cu, and Ru), CeTIn$_5$ (T=Co, Rh, and Ir)
and pressure-induced quantum critical point in CeRhIn$_5$
as studied in our previous works,
Ref.~\onlinecite{mm_2009},~\onlinecite{mm_2010},
and~\onlinecite{mm_2019}, respectively.
Thus the working computational setup has been applied to elucidate the magnetism
trends around QCP as precisely as have been done for other representative HF compounds.
In practice, we define $J_{\rm K}$ at $J_{\rm Hund}=0$ as $J_{\rm K,0}$
and then sweep a multiplicative factor $\alpha = J_{\rm K}/J_{{\rm K},0}$,
calculating the temperature dependence of staggered magnetic susceptibility $\chi(\pi,T)$
for each $\alpha$.
In this way we can see
where in the neighborhood of the QCP our target material
with $\alpha$ corresponding to the realistic number, $J_{\rm Hund}=1$~eV,
resides on the Doniach phase diagram.

We calculate the staggered
magnetic susceptibility $\chi(\pi)$
with the two-particle Green's function
following the formalism developed in Ref.~\onlinecite{otsuki_2009_II}
and using
a random-dispersion approximation to decouple it into single-particle
Green's functions~\cite{gebhard_1997}
which would enhance the transition temperature,
in addition to the single-site mean-field nature in DMFT.
\begin{figure}
\scalebox{0.7}{
\includegraphics{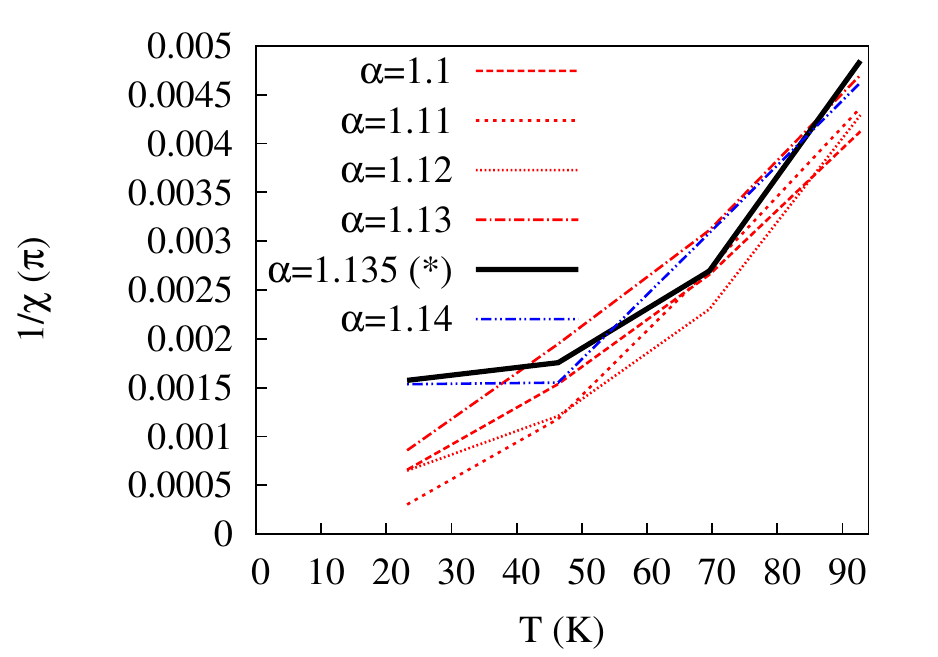}}
\caption{\label{fig::ssus_inv} (Color online) Calculated temperature dependence of
  the reciprocal of
  staggered magnetic susceptibility for CeCu$_6$, the reference compound.
  The data with $\alpha=1.135$, specified with an asterisk, corresponds to the realistic data point.}
\end{figure}
The calculated data for $1/\chi(\pi)$
is shown in Fig.~\ref{fig::ssus_inv} for the case of CeCu$_6$.
The temperature dependence of the reciprocal of
the staggered magnetic susceptibility $1/\chi(\pi)$
is observed for each $J_{\rm K}=\alpha J_{{\rm K},0}$ and we extrapolate it linearly to the low temperature
region to see if there is a finite N\'{e}el temperature.
We identify
that the N\'{e}el temperature vanishes in the
parameter range $1.13J_{{\rm K},0}<J_{\rm K}<1.135J_{{\rm K},0}$, where $J_{{\rm K},0}$
is the Kondo coupling at $J_{\rm Hund}=0$.
The realistic data point
is obtained by plugging in $J_{\rm Hund}=1$~(eV)~\cite{mm_2009}
and $\epsilon_{f}=-1.61$~(eV)
(as can be found in Table~\ref{table::e_f}) to Eq.~(\ref{eq::sw}) to be $J_{\rm K}=1.1347J_{\rm K,0}$.
Thus the data in Fig.~\ref{fig::ssus_inv} shows that CeCu$_6$
is almost right on the magnetic QCP where the N\'{e}el temperature disappears.

The same procedures are applied to
all other target materials.

\section{Results}
\label{sec::results}

\begin{figure}
\scalebox{0.7}{
\includegraphics{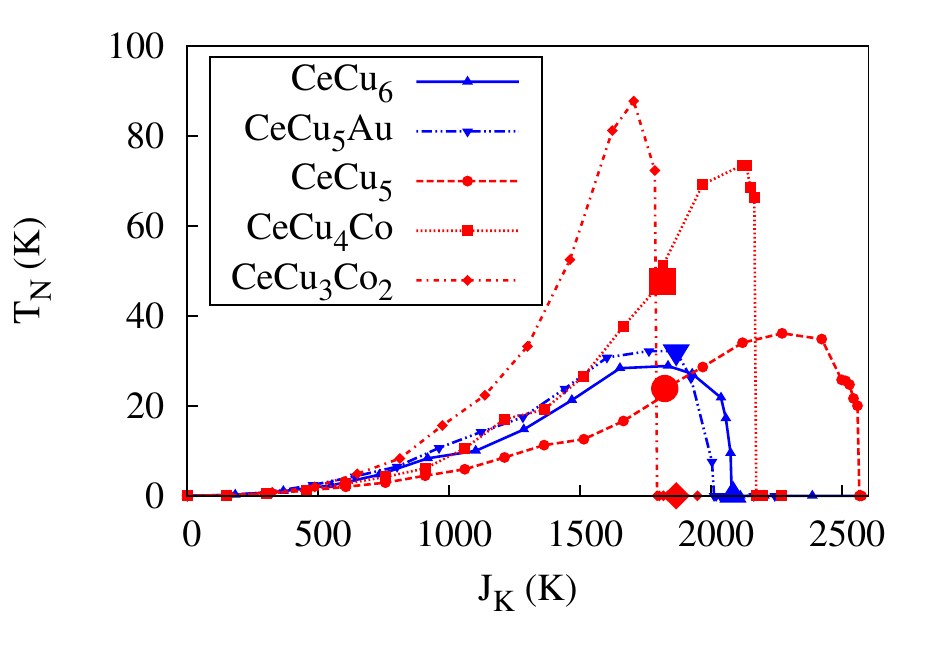}}
\caption{\label{fig::Doniach} (Color online) Realistic Doniach phase diagram for the target compounds
with the bare energy scale of the Kondo couplings.}
\end{figure}

Plotting calculated N\'{e}el temperatures with respect to $J_{\rm K}=\alpha J_{{\rm K},0}$,
the Doniach phase diagram is constructed for each target material as shown in Fig.~\ref{fig::Doniach}.

By rescaling the horizontal axis of the Doniach phase diagram as follows,
$t \equiv (J_{\rm K}-J_{\rm K,QCP})/J_{\rm K,QCP}$
to inspect the dimensionless distance to the QCP independently of the materials~\cite{mm_2009, mm_2010},
we end up with the main results as shown in Fig.~\ref{fig::rescaled_Doniach}.

\subsection{CeCu$_6$ vs CeCu$_5$}

Remarkably, CeCu$_6$ falls almost right on top of magnetic QCP
in Fig.~\ref{fig::Doniach}. Also it is seen that
the energy scales for antiferromagnetic order
are on the same scale for CeCu$_6$ and CeCu$_5$
as seen in the vertical-axis
scales for the calculated N\'{e}el temperatures.
This may be reasonable considering the similar chemical composition
between CeCu$_6$ and CeCu$_5$.

Here we note that overestimates of the calculated N\'{e}el temperature
are unavoidable due to the single-site nature of DMFT
and approximations involved in the estimation of two-particle Green's function~\cite{mm_2009}.
Thus the calculated N\'{e}el temperature for CeCu$_5$ falling in the range of 20K
should be compared to the experimental value of 4K~\cite{bauer_1987,bauer_1991}
only semi-quantitatively.
Nevertheless, expecting that the same degree of systematic deviations are
present
in all of the data for the target compounds, we can safely inspect the relative trends between CeCu$_6$ and CeCu$_5$.

\subsection{Magnetic QCP in Au-doped CeCu$_6$}
\label{sec::results::cecu6}

In Fig.~\ref{fig::Doniach} it is seen
that doping Au into CeCu$_6$ only slightly
shifts the energy scales competing between magnetic ordering and Kondo screening.
Most importantly, Au-doping drives the material towards the antiferromagnetic phase
and the magnetic QCP is identified in the region Ce(Cu$_{1-\epsilon}$Au$_{\epsilon}$)$_6$ with $\epsilon\ll 1$,
which is consistent with the experimental trends of magnetism~\cite{1994_cecu5.9au0.1,2000_cecu5au}.
This has been achieved with the control of an effective degeneracy of orbitals
incorporating the realistic width of level splittings in the localized $4f$-orbital,
putting the characteristic energy scales in magnetism under good numerical
control in the present modeling.

While quantitative success for Ce(Cu$_{1-\epsilon}$Au$_{\epsilon}$)$_6$ ($0\le\epsilon\ll 1$)
concerning the location of the magnetic QCP is seen,
some qualitative issues may be considered to be
on the way to address magnetic quantum criticality, since this particular materials family
represents the local quantum criticality scenario~\cite{si_2001,coleman_2001,rmp_2020}
where a sudden breakdown of the Kondo effect is discussed to occur on the basis
of a Kondo-Heisenberg model. In our realistic model, the
exchange interaction between localized $4f$-electrons naturally come in
as a second-order perturbation process
with respect to the Kondo coupling~\cite{yosida_1957},
which is the RKKY interaction~\cite{rk_1956,kasuya_1956,yosida_1957}.

Even though
no special place for another Heisenberg term is identified in our realistic Kondo lattice model,
there should indeed be other terms that are not explicitly considered: for example,
with very well localized $4f$-electrons, there is another indirect exchange coupling~\cite{campbell_1972}
that work via the following two-steps: (a) intra-atomic exchange coupling between $4f$-spin and $5d$-spin
and (b) inter-atomic hybridization between $5d$-band and other conduction band. Notably, in this channel
the coupling between $5d$ and $4f$ is ferromagnetic, which is in principle in competition
against the antiferromagnetic Kondo coupling that we mainly consider here.

For REPM compounds
such as Nd-Fe intermetallics, the latter indirect exchange coupling,
which we denote $J_{\rm RT}$ for the convenience
of reference as the effective coupling between rare-earth elements
and transition metals,
is dominant because $4f$-electrons are even more well localized than in Ce$^{3+}$-based compounds.
There the Kondo couplings
are not in operation practically, since $f$-$c$ hybridization is weak
and Kondo couplings are at too small energy scales as compared to other exchange couplings.
Now that we bring HF materials and REPM compounds on the same playground,
the $f$-$d$ indirect exchange couplings should also have been given more attention even though
there is at the moment only some restricted prescriptions~\cite{mm_2016,toga_2016}
to downfold a realistic number into $J_{\rm RT}$.

This indirect exchange coupling can motivate
the Heisenberg term on top of the realistic Kondo lattice model, even though it is to be noted
that the sign of such extra Heisenberg terms is ferromagnetic. This may pave the way to define
a realistic version of the Kondo-Heisenberg model~\cite{si_2001,coleman_2001,rmp_2020}. Since $J_{\rm RT}$'s can compete against RKKY
at most only on the same order, the presence of the $J_{\rm RT}$ terms would not significantly alter
the position of magnetic QCP, which is brought about by the Kondo coupling that competes against RKKY
as a function of exponential of the reciprocal of the coupling constants. This way, it is hoped
that there might be a way to reconcile the local QCP scenario for Au-doped
CeCu$_6$
and the present realistic modeling for the magnetic QCP
focusing on the characteristic energy scales involving the Kondo effect.

Recent time-resolved measurements and theoretical analyses based on DMFT~\cite{NatPhys2018,PRL2019}
for Ce(Cu,Au)$_6$ also provides a way to reconcile the local QCP scenario
and experimentally detected signals from the possible Kondo quasiparticles
on the real-time axis within the non-crossing approximation (NCA)~\cite{kuramoto_1983,bickers_1987}
as the impurity solver in DMFT.
Since our DMFT results are based on quantum Monte Carlo (QMC)
formulated on the imaginary-time axis, migrating to the real-time data
via analytic continuation poses a challenging problem~\cite{jarrel_gubernatis_1996,otsuki_2017},
while the solution of the quantum many-body problem is numerically
exact with QMC. Thus the location of QCP derived from static observables
would be better addressed with the present framework.

Still our numerically exact solution is limited to the imaginary-time direction
and effects of the real-space fluctuations are not incorporated
in the single-site DMFT. Recently, theoretical comparison between
an exact solution of the lattice problem and DMFT has been done~\cite{PRL20190606}
and an artifact of DMFT to overestimate the region of antiferromagnetic phase
has been demonstrated. In this respect, the present location of magnetic QCP
right below CeCu$_6$ should also reflect the same artifact: if the spatial fluctuations
are properly accounted for, the magnetic phase would shrink and the position
of CeCu$_6$ would shift slightly toward the paramagnetic side.

\subsection{QCP to which CeCu$_5$ is driven by Co-doping}
\label{sec::CeCuCo5}

Co-doping in CeCu$_5$ shifts the energy scales more strongly than seen
in Au-doped CeCu$_6$. It is seen in Fig.~\ref{fig::Doniach} that the QCP is
driven toward the smaller $J_{\rm K}$ side, reflecting
the underlying physics that
Kondo-screening energy scale is enhanced as Co replaces Cu.
The origin of the enhanced Kondo screening is seen in Fig.~\ref{fig::hyb}
where anomalous peaks below the Fermi level are coming in which should
come from the almost ferromagnetic conduction band
which grows into the ferromagnetism in the Co-rich side of the composition space
in Ce(Cu,Co)$_5$.
With 40\% of Co, the $4f$-electron QCP is already passed
and CeCu$_3$Co$_2$ resides in the Kondo-screened phase.
Thus it is found that the magnetic QCP of Ce(Cu$_{1-x}$Co$_{x}$)$_5$
is located in $0.2<x_{\rm c}<0.4$.
We note that the crystal structure and crystal-field splitting
have been fixed to be that of the host material CeCu$_5$.
In reality, the QCP may be encountered with smaller Co concentration.

In the present simulations, we have neglected the possible ferromagnetism in the ground state
contributed by the $3d$-electrons in Co.
Referring to the past experiments for Ce(Cu,Co)$_5$ described
in Ref.~\onlinecite{girodin_1985},
the absence of the observed Curie temperatures
for the Cu-rich side with the concentration of Cu beyond 60\%
in the low-temperature region seems to be consistent with our computational setup
in the present simulations. Even though
other past work~\cite{lectard_1994} for an analogous materials family Sm(Co,Cu)$_5$ does show
residual Curie temperature in the Cu-rich region,
it should be noted that there is a qualitative difference
in the nature of the conduction band of Cu-rich materials for the Sm and Ce-based families.

\subsection{Universal and contrasting trends}

Co-doped CeCu$_5$ and Au-doped CeCu$_6$ represent the different mechanisms
where Co enhances $f$-$d$ hybridization with the $3d$-electron magnetic fluctuations in the conduction electrons,
while Au rather weakens $f$-$d$ hybridization, being without $d$-electron magnetic fluctuations.

The trend in magnetism
comes from the relative strength
of exchange coupling
between localized $4f$-electron and delocalized conduction electrons.
Among Ce-based intermetallic compounds,
a general trend in the hybridization $\Delta$ is seen to be like the following
\begin{equation}
J_{\rm K}(\mbox{Ce-Au})<J_{\rm K}(\mbox{Ce-Cu})<J_{\rm K}(\mbox{Ce-Co})
\label{eq::hyb_trend}
\end{equation}
as is partly seen in Ref.~\onlinecite{mydosh_1993} for other materials family
Ce$T_2$Si$_2$ ($T$=transition metals) - somewhere in the sequence of the trend
written schematically
in Eq.~(\ref{eq::hyb_trend}),
a magnetic quantum critical point between antiferromagnetism located on the relatively
left-hand side and paramagnetism located on the relatively right-hand side is encountered
within the range where $3d$-electron ferromagnetism from Co does not dominate.
The overall one-way trend from antiferromagnetism on the left-most-hand side
to paramagnetism on the right-most-hand side in Eq.~(\ref{eq::hyb_trend}) is universal
around the magnetic quantum criticality, while the contrasting trend
between Au-doped case and Co-doped case in Ce-Cu intermetallics
is seen from
the position of the Ce-Cu intermetallics
concerning the directions toward which the dopant elements drive.

The opposing trends coming from $3d$-metal dopant and $5d$-metal dopant might help in implementing
a fine-tuning of the material in a desired proximity to QCP in a possible materials design
for REPM's
as discussed below in Sec.~\ref{sec::md}.

\section{Discussions}
\label{sec::discussions}

\subsection{Validity range of the Kondo lattice model}
While we have defined the Kondo lattice model
referring to the electronic structure
of the target materials, the limitations on the validity range
of such downfolding approach~\cite{mm_2009,imada_2010}
should be kept in mind
in assessing the implications
of the present results. In Sec.~\ref{sec::results::cecu6},
we have already discussed the possible relation of our model
to the Kondo-Heisenberg model that has been extensively used in the local QCP scenario~\cite{si_2001,coleman_2001,rmp_2020}
for Ce(Cu$_{1-\epsilon}$Au$_{\epsilon}$)$_6$. In a wider context,
the spirit of
the so-called $s$-$d$ exchange model that was originally introduced by Vonsovskii~\cite{vonsovskii_1946} and
Zener~\cite{zener_1951} in the early days
of the theory of ferromagnetism is still alive in the indirect exchange coupling $J_{\rm RT}$.
This has been dropped in the present modeling for Ce-based compounds. Here we have assumed that
the localization of the $4f$-electron in our Ce$^{3+}$-based compounds is good enough to assure the applicability
of the Kondo lattice model:
at the same time, it is presumed that
our $4f$-electrons in Ce$^{3+}$-based $f$-$d$ intermetallics are not so well localized as are the case in Pr$^{3+}$ or Nd$^{3+}$-based
$f$-$d$ intermetallics.
This means that the Kondo coupling coming from $f$-$c$ hybridization would dominate over
the $J_{\rm RT}$'s coming from the indirect exchange coupling~\cite{campbell_1972}. Such subtle interplay between different
exchange mechanisms
can depend on the material. Since we did not address $J_{\rm RT}$ in the present studies, the outcome of the possibly
competing
exchange interactions is not included in the present scope.
Possible subtle aspects coming from the local QCP scenario might reside in this particular leftover region.
If one would further opt for an alternative scenario~\cite{sluchanko_2015},
it may be useful to further investigate the effect of these
dropped terms, and include them through improved algorithms based on better intuition.
Although a completely {\it ab initio} description is desirable,
to make the problem tractable we are forced into making some approximations.
Here it is at least postulated that the validity of the relative
location of magnetic QCP can be assured in the present description
because we have put the most sensitive coupling channel, Kondo physics, under good numerical control.

A few more discussions on the validity range of the Kondo lattice model
and possible extensions are in order:

\subsubsection{Toward more unbiased downfolding}
The terms in our low-energy effective models have been defined
targeting at the particular physics, namely, the RKKY interaction and Kondo physics.
While this strategy has been good enough to address the relative trends among the target materials
around the magnetic QCP, it may well have happened that other relevant terms have been dropped
that do not significantly affect the location of QCP. In this regard
it may be preferred either
a) to downfold from the realistic electronic structure
to the low-energy effective models in a more unbiased way, at least proposing
all possible candidate terms and eliminating some of them only in the final stage
according to a transparent criterion e.g. referring to the relevant energy window
or b) to work on the observables directly from first principles without downfolding.
While b) does not look very feasible, a) might pose a feasibly challenging problem
with a possible help from machine learning~\cite{rmp_2019} in systematically classifying the candidate terms
even for such materials with multiple sublattices, multiple orbitals and relatively large
number of orbital degeneracy as imposed from $d$-electrons and $f$-electrons.

\subsubsection{Effects of valence fluctuations}
Valence fluctuations have not been entirely incorporated in the present description
of Ce compounds. Other scenario for Au-doped CeCu$_6$ that emphasizes
the relevance of valence fluctuations are recently discussed~\cite{2018}.
We have described at least the magnetism trends around the QCP in CeCu$_6$ and CeCu$_5$Au
only with localized $4f$-electrons. Apparently valence fluctuations may not be dominant
at least for magnetism. We can restore the charge degrees of freedom for $4f$-electrons
and run an analogous set of
simulations for a realistic Anderson lattice model in order
to see any qualitative difference comes up on top of the localized $4f$-electron physics.
Often the typical valence states for Ce, Ce$^{4+}$ or Ce$^{3+}$, are not so clearly distinguished:
even in the present Kondo lattice description, $(4f)^0$ state with Ce$^{4+}$
are virtually involved in the Kondo coupling
and localized $4f$-electrons
even contribute to the Fermi surface~\cite{otsuki_2009_FS}.
To pick up a few more cases, for actinides
or $\alpha$-Ce, one can either discuss on the basis of localized $f$-electrons
and define the Kondo screening energy scale spanning up to 1000K,
or convincing arguments can be done
also on the basis of delocalized $4f$-electrons
emphasizing the major roles played by valence fluctuations.
Given that it does not seem
quite clear how precisely
the relevance or irrelevance of valence fluctuations should be formulated
for the description of magnetism trends, here we would claim only the relative simplicity
of our description for magnetic QCP in Ce(Cu$_{1-\epsilon}$Au$_{\epsilon}$)$_6$ ($\epsilon\ll 1$).
This simplification may well come with the restricted validity range.

\subsection{Implications on the coercivity of REPM}
\label{sec::md}

Observing that the magnetic
QCP can be encountered in the chemical composition space of Ce(Cu,Co)$_5$,
we note that slowing down of spin dynamics when the system crosses over to QCP
can be exploited in intrinsically blocking the magnetization reversal processes
in REPM to help the coercivity. Since coercivity is a macroscopic and off-equilibrium
notion, it is still much under development to formulate a theoretical
  bridge over the gap between the microscopic equilibrium
properties and macroscopic coercivity. At least with QCP, diverging length scales of fluctuations
and diverging relaxation times can in principle reach the macroscopically relevant
spatial and time scales to help coercivity. Range of
the critical region on the temperature axis and on the composition space
would depend on each specific case.

In Sm-Co magnets,
even though it is clear that the cell boundary phase
intrinsically carries the coercivity~\cite{nesbitt,ohashi_2012,navid_2017},
precise characterization
of the inter-relation among the intrinsic properties, microstructure, and coercivity has been
still
under investigation~\cite{ohashi_2012,navid_2017,chris_2018}.
Since Sm(Cu,Co)$_5$ can be considered as a hole analogue of Ce(Cu,Co)$_5$
in the lowest $j=5/2$ multiplet of Ce$^{3+}$,
with a quest for QCP both for magnetism and possibly also for valence fluctuations,
it may help to consider the possible role of QCP in Sm(Cu,Co)$_5$ for the intrinsic part of the coercivity mechanism.
Considering the electron-hole analogy, possible effect from QCP for Sm(Cu$_{1-x'}$Co$_{x'}$)$_5$
can be expected in the concentration range $x'_{\rm c}\simeq 1-x_{\rm c}$
(here $x_{\rm c}$ is defined in Sec.~\ref{sec::CeCuCo5})
which fall in $0.8>x'_{\rm c}>0.6$.
This may be compared favorably
with the experimentally discussed~\cite{nesbitt} concentration of Cu
in Sm-Co magnets, where up to around 35\% of Cu
in the cell-boundary phase made
of Sm(Co,Cu)$_5$, especially in the triple-junction area~\cite{zhang_2000,xiong_2004},
has been correlated with the emergence of good coercivity.

\section{Conclusions and outlook}
\label{sec::conc}

Realistic modeling for Au-doped CeCu$_6$ and Co-doped CeCu$_5$
successfully describes the trends in magnetism involving QCP
on the basis of the localized $4f$-electrons. One of the archetypical HF materials
family, CeCu$_6$, and its Au-doping-induced QCP can be described
within a magnetic mechanism with the terms
that can be naturally downfolded from the realistic electronic structure
in the spirit of the Anderson model~\cite{anderson_1961},
without explicitly invoking valence fluctuations or introducing additional Heisenberg terms.
We believe we have just put the characteristic energy scales of the target materials
around QCP under good numerical control
in having succeeded in addressing the relative trends in magnetism around QCP.
We do not rule out other subtlety around QCP that may come from other terms~\cite{si_2001,coleman_2001,rmp_2020}
that are not included
in the present simulation framework. As long as the dominating energy scales
are concerned, those other terms
would not significantly alter the magnetism trend around QCP.

Co-doping in CeCu$_5$ drives the material on a wider scale
on the chemical composition axis as compared to Au-doped CeCu$_6$.
This is caused by magnetic fluctuations
in the paramagnetic conduction band that is on the verge of ferromagnetism.
For the $4f$-$3d$ intermetallic paramagnets in REPM in general,
small changes in the $3d$-metal concentration
can drive the material around
in the proximity of quantum criticality on the chemical composition space,
rendering it easy
to encounter critical regions in a microstructure with an appropriate spatial
variance in the microchemistry.

Ce(Co,Cu)$_5$ represents one of the earliest and most typical materials family
in REPM~\cite{rmp_1991, li_and_coey_1991}. The lattice structure
of the materials family RT$_5$ including Ce(Co,Cu)$_5$
can be transformed into R$_2$T$_{17}$
and RT$_{12}$~\cite{li_and_coey_1991} (R=rare earth and T=Fe group elements),
and a local structure
around the rare-earth sites in the champion magnet compound
R$_2$Fe$_{14}$B (R=rare earth) resembles RT$_5$
as described in Sec.~III~A of Ref.~\onlinecite{rmp_1991}.
With our results for Ce(Cu,Co)$_5$ in relation to Ce(Cu,Au)$_6$ concerning QCP,
it has been suggested that potentially various properties of derived compounds from the
RT$_5$ archetypical series~\cite{chris_2019}
residing in REPM, especially physics in the crossover to QCP,
can be exploited for the possible intrinsic contribution to coercivity.

\begin{acknowledgments}
  MM's work in Institute for Solid State Physics (ISSP), University~of~Tokyo
  and High Energy Accelerator Research Organization (KEK)
  is supported by Toyota~Motor~Corporation.
  The author thanks C.~E.~Patrick for his careful reading of the manuscript.
  The author benefited from comments given by
  J.~Otsuki and H.~Sepehri-Amin.
  The author gratefully acknowledges
  helpful discussions with H.~Shishido, T.~Ueno, K.~Saito
  in related projects,
  crucial suggestions given by T.~Akiya
  pointing to the particular materials family R(Cu,Co)$_5$ (R=rare earth),
  and an informative lecture given by K.~Hono in the early stage of this project.
  The present work was partly supported by JSPS~KAKENHI~Grant~No.~15K13525.
  Numerical computations were executed on
  ISSP Supercomputer Center, University of Tokyo
  and Numerical Materials Simulator in National Institute for Materials Science.
\end{acknowledgments}

\end{document}